\documentclass[12pt,letterpaper]{article}
\usepackage[latin1] {inputenc} 
\usepackage{amsmath} \usepackage{amsfonts} \usepackage{amssymb} 

\newcommand{\be}{\begin{equation}}
\newcommand{\eq}{\end{equation}}
\newcommand{\bea}{\begin{eqnarray}}
\newcommand{\eqa}{\end{eqnarray}}
\begin{document}

\title{BFT embedding of the Green-Schwarz superstring and the pure spinor formalism}
\author{Alejandro Gaona\footnote{email:gaona@nucleares.unam.mx}\,\, and
 J. Antonio Garc\'\i a\footnote{email:
garcia@nucleares.unam.mx} \\
\em Instituto de Ciencias Nucleares, \\
\em Univesidad Nacional Aut\'onoma de M\'exico\\
\em Apartado Postal 70-543, M\'exico D.F., M\'exico}
\maketitle
\abstract{ We worked out the Batalin-Fradkin-Tyutin (BFT) conversion program of second
class constraints to first class constraints in the GS superstring using light
cone coordinates. By applying this systematic procedure we were able
to obtain a gauge system that is equivalent to the recent model proposed in
\cite{BM} to relate the GS superstring to the pure spinor formalism.}

\section{Introduction}

The covariant quantization of GS superstring is an important open problem for
string theory and for the general theory of constrained systems.  The problem
is as old as the initial proposal of the classical action by Green and
Schwarz \cite{GS}, and
many different ways to tackle it with a wide spectrum of techniques was
worked along the years. Among them are covariant conversion
procedures (with an infinite number of auxiliary fields), 
BRST program with infinitely reducible constraints \cite{CC}, the use of
light cone coordinates and the question of conformal  invariance \cite{LILC}, 
and recently the pure spinor 
formalism \cite{PS}, gauging cosets \cite{GC}, and BRST extensions attempting to lift the bosonic
spinor constraint of the pure spinor formalism by introducing more ghost variables 
but a finite number of them \cite{LPS}. 

The problem with the quantization of the GS superstring lies in the fact that 
we do not know
a procedure to separate the first and second class fermionic constraints in a
manifestly Lorentz invariant way. The first class sector of these fermionic
constraints is responsible for the $\kappa$ symmetry of the superstring while
the second class constraints appear as in any other fermionic system because
the Lagrangian is linear in the time derivatives of the fermionic variables.
So we face the problem of trying to quantize
the system without splitting the constraints or split them in a non-covariant
way and use a somewhat complicated Dirac bracket to perform the quantization. The
recent proposed pure spinor formalism is an important step in the direction of
a covariant quantization program for the GS superstring. This formulation is 
covariant but the price to pay is a radical departure from the BRST standard
techniques, introducing a constraint in the bosonic ghost sector of the
theory, know as the pure spinor constraint. The idea could be
related to the use
of all the constraints (not only the first class constraints) to construct the BRST
operator. The condition $Q^{2}=0$ implies then a constraint in the ghost sector.
Nevertheless,
this quantization program has been proved to be very useful in many
calculations that do not imply the explicit solution of the constrained ghost
relation. The other standard formalism to describe
the supersymmetric string, the RNS model, has
the supersymmetry realized in the worldsheet so the space time
supersymmetry is not manifest. As a consequence of this fact the
spectrum of the RNS theory is not Lorentz covariant. We need to impose a projection of
states in its spectrum, the so called GSO projection to recover a Lorentz invariant spectrum. 
Without a manifest space
time supersymmetric action is very difficult -if not impossible- to describe the superstring
 in Ramond-Ramond backgrounds.

In what follows we will present a solution to the long standing 
problem of
the implementation of the conversion program of second class constraints
into first class constraints by adding to the GS action new fermionic variables with a standard
symplectic structure. Our result is
completely equivalent to the recent model proposed by Berkovits and Marchioro
(BM) to relate the GS superstring with the pure spinor formalism \cite{BM}.
From our point of view, the  key observation given by these authors is that in 
the extended phase space the Lorentz generators
close in the corresponding Lorentz algebra up to and exact BRST term. Moreover, and in spite 
of the non Lorentz covariant approach, the
quantization program can be implemented because the anomaly that comes from
the nonlinear products of the new added variables can be canceled using standard
BRST techniques. A previous attempt to relate the pure spinor formalism with the GS superstring
using standard BRST 
approach is \cite{OT} where the authors were able to modify the GS original action
with a non-local term, that when added to
the GS action, render it BRST invariant. 

In our approach we do not need to fix the gauge (the semi-light cone gauge)
nor to add variables other than the ones required by the conversion procedure.
Our work is in close relation with a recent proposed action given in \cite{AK}
where the authors start from a modified GS action doubling the fermion sector
and introducing an interaction between the fermionic sectors by hand. These
authors claim that theirs modified GS action is equivalent to the BM action by fixing the
gauge and performing a Darboux transformation that changes the remaining Dirac
bracket to a standard symplectic structure.
We will see that this model can be explained and simplified using the
conversion approach presented here. 

In  section 2 we will review the basic ideas of the BFT program. Section 3
will be used to expose our main results and the GS gauged action. In section 4 we present
some comments on the relation between our results and the Aisaka-Kazama (AK) action 
\cite{AK}, and  in section 5 the conclusion and some notes about possible future
work.

\section{Birds eye to BFT procedure}

The aim of the BFT method \cite{BFT} is to develop a systematic approach to the conversion
of a general set of constraints into an algebra of first class constraints by
adding to the original phase space an appropriate  number of new variables with its own 
symplectic structure. The method provides us with a procedure for the
conversion of second class constraints into first class ones in this extended
phase space and a procedure to modify any observable, including any previous 
first class constraint in such a way that we can construct an effective first
class gauge algebra and an effective action consistent with the whole conversion
procedure. It is based on homological perturbation theory and is, in this respect,
very similar to the iterative construction of the BRST charge given a gauge
algebra.

Suppose that we have a set of constraints  where some constraints are second
class 
$\chi_\alpha$, $\{\chi_\alpha,\chi_\beta\}=C_{\alpha\beta}$ and some are first
class $\phi_m$ in a phase space defined by the coordinates $z^i$ with the standard
symplectic form $\sigma^{ij}$. Now we will add $\xi_\alpha$ new variables with
symplectic structure $\omega^{\alpha\beta}$ to the original phase space.
 The idea is to construct a new set
of constraints $\tilde \chi_\alpha$ satisfying the algebra
\be
\label{BFT-cond}
\{\tilde\chi_\alpha,\tilde\chi_\beta\}=0.
\eq 
To solve for $\tilde\chi_\alpha$ we propose a solution in power series of
the new variables
\be
\tilde\chi(z,\xi)=\sum_n X_{\alpha}^{(n)}, 
\eq 
where the $n=0$ term coincides with the original constraint $\chi_\alpha$
and  $X^{(n)}_\alpha$ is a term proportional to $\xi^n$ in the series
expansion. The solution, up to a canonical transformation, in the extended phase
space is \cite{BFT}
 \be X^{(0)}_\alpha=\chi_\alpha, \quad 
X^{(1)}_{\alpha}=X_{\alpha\gamma}\xi^\gamma,\quad X_{\alpha\gamma}
\omega^{\gamma\delta}X_{\beta\delta}=-C_{\alpha\beta}(z),
\eq 
and for the next terms $n\geq 2$ in the power series
\be
X^{(n+1)}_\alpha=-\frac{1}{n+2}\xi^\beta\omega_{\beta\gamma}X^{\gamma%
\rho}X^{(n)}_{\rho\alpha},
\eq 
where
\be X^{(1)}_{\alpha\beta}=\{\chi_{[\alpha},X^{(1)}_{\beta]}\}, \quad
X^{(n)}_{\alpha\beta}=\sum_{m=0}^n\{X^{(n-m)}_\alpha,X^{(m)}_\beta\}+
\sum_{m=0}^{n-2}\{X^{(n-m)}_\alpha,X^{(m+2)}_\beta\}_\xi,
\eq 
and the first bracket is evaluated using only the original phase space variables 
and the second using only the new variables.

The same idea works also to extend any function $f(z)$  of the original variables
$z$ to a new function $\tilde f(z,\xi)$ as a solution in power series of the
new variables $\xi$. This series must satisfy the condition
\be
\{\tilde\chi_\alpha, \tilde f\}=0, \quad \tilde f=\sum_n F^{(n)},
\eq 
where $F^{(0)}=\tilde f(z,0)=f(z)$ and $F^{(n)}$ is the term proportional to
$\xi^n$. The solution is \cite{BFT}
\be
F^{(n+1)}=-\frac{1}{n+1}\xi^\beta\omega_{\beta\gamma}
X^{\gamma\rho}F^{(n)}_{\rho},
\eq 
where
\be 
F^{(0)}_{\alpha}=\{\chi_{\alpha}, f(z)\}, \quad
F^{(1)}_{\alpha}=\{X^{(1)}_{\alpha}, f(z)\}+\{\chi_{\alpha},
F^{(1)}\}+\{X^{(2)}_\alpha, F^{(1)}\}_\xi,
\eq
and
\be F^{(n)}_{\alpha}=\sum_{m=0}^n\{X^{(n-m)}_\alpha,F^{(m)}\}+
\sum_{m=0}^{n-2}\{X^{(n-m)}_\alpha,F^{(m+2)}\}_\xi+\{X^{(n+1)}_{\alpha},
F^{(1)}\}_\xi.
\eq 
In particular, we can extend the original first class constraints $\phi_m$ to a
new set of constraints $\tilde \phi_m$ in such a way that all the new constrains close in
a new gauge algebra in the extended phase space. In what follows we will need only
terms up to second order in the new variables.

An interesting corollary of the conversion approach is that the original Dirac
bracket can be recovered using
\be
\{\tilde A,\tilde B\}|_{\xi=0}=\{A,B\}_{D},
\eq
for any two functions of the original phase space $A,B$ that were extended to
$\tilde A, \tilde B$ as can be easily checked.

\section{BFT embedding of the GS superstring}

We start from the GS action that we write in the form
\be
\label{GS-action}
S=-\frac12 \int d^{2}\zeta \Big[\sqrt{-g}g^{ij}\Pi^{\mu}_{i}\Pi_{\mu
j}+2\varepsilon^{ij}\Pi^{\mu}_{i}(W^{1}_{j\mu}-W^{2}_{{j\mu}})-2
\varepsilon^{ij} W^{1\mu}_{i}W^{2}_{j\mu}\Big],
\eq
where
\be
W^{A\mu}_{i}=i\theta^{A}\gamma^{\mu}\partial_{i}\theta^{A}, \quad 
\Pi^{\mu}_{i}=\partial_{i}x^{\mu}-\sum_{A}W^{A\mu}_{i}.
\eq
As usual the bosonic constraints can be obtained by setting to zero  the energy-momentum
tensor. Taking the conformal gauge, the first order
Lagrangian associated is 
\be
{\cal L}=\dot x^{\mu}p_{\mu}+\dot\theta^{A}_{\alpha}p^{A}_{\alpha}-H_{c}-
\lambda^{A}_{\alpha} d^{A}_{\alpha},
\eq
where
\be
p_{\mu}=\Pi_{0\mu}-(W^{1}_{1\mu}-W^{2}_{1\mu}),
\eq
and
\be
d^{1}_{\alpha}=p^{1}_{\alpha}-i(\theta^{1}\gamma^{\mu})_{\alpha}
(p_{\mu}-x'_{\mu}+W^{1}_{1\mu}),
\eq
\be
d^{2}_{\alpha}=p^{2}_{\alpha}-i(\theta^{2}\gamma^{\mu})_{\alpha}
(p_{\mu}+x'_{\mu}-W^{2}_{1\mu}),
\eq
are the fermionic constraints. The canonical Hamiltonian is
\be
H_{c}=\frac12\Big[\Big(p_{\mu}+W^{1}_{1\mu}-W^{2}_{1\mu}\Big)^{2}+
\Big(x'_{\mu}-\sum_{A}W^{A}_{1\mu}\Big)^{2}\Big]=
\frac12\Big(\Pi_{0}^{2}+\Pi_{1}^{2}\Big).
\eq
The bosonic constraints 
\be
{\cal H}=\frac12\Big(\Pi_{0}^{2}+\Pi_{1}^{2}\Big),\quad 
{\cal H}_{1}=\Pi_{0}^{\mu}\Pi^{\mu}_{1},
\eq
can be written in the form
\be
\hat T={\cal H} + {\cal H}_{1}=\frac12 \hat\Pi^{2},\quad 
T={\cal H} - {\cal H}_{1}= \frac12 \Pi^{2},
\eq
where
\be
\label{Pi-mu}
\Pi_{\mu}=p_{\mu}-x'_{\mu}+2W^{1}_{1\mu},\quad 
{\hat \Pi}_{\mu}=p_{\mu}+x'_{\mu}-2W^{2}_{1\mu}.
\eq
The algebra of constraints naturally splits into two sectors 
\be
\{d_{1\alpha},d_{1\beta}\}=2i\gamma^{\mu}_{\alpha\beta}\Pi_{\mu},\quad
\{d_{2\alpha},d_{2\beta}\}=2i\gamma^{\mu}_{\alpha\beta}\hat\Pi_{\mu},
\eq
\be
\{d_{1\alpha},d_{2\beta}\}=0,\quad 
\{d_{1\alpha},\hat \Pi_{\mu}\}= 0,\quad
 \{d_{2\alpha},\Pi_{\mu}\}=0, \quad   \{\Pi_{\mu},\hat\Pi_{\nu}\}=0.
\eq
To separate the constraints into first and second class we will write them in
the light cone coordinates and divide the $\alpha,\beta$ spinor indices using
the spinorial representation of the little group SO(8). The result is that
the constraints $d^{A}_{a}=0$,  ($\Pi^{+}\not=0$) are second class while the
rest of the constraints $d^{A}_{\dot a}=0, T=0,\hat T=0$ are first class. This
fact allow us to count the number of degrees
of freedom for the superstring given us the correct result, as expected. 
In what follows we will not need the details of
this first class algebra as our aim is to construct a new effective gauge
algebra. To that end we
extend the original phase space
$x^{\mu},p_{\mu},\theta^{A}_{\alpha},p^{A}_{\alpha}$ by adding the fermionic 
variables $S_{a}$ with the symplectic structure\footnote{In what follows we will work on the
sector with index 1 and for simplicity we will remove this index from our
equations. The other sector can be worked in the same way.}.
\be
\{S_{a},S_{b}\}=i\delta_{ab}.
\eq
Searching a solution for the condition (\ref{BFT-cond}) 
\be
\{\tilde d_{a}, \tilde d_{a}\}=0,
\eq
in power series of $S$ give us a very simple result. The solution is linear in $S$ and yields
\be
\label{new-c-1}
\tilde d_{a}=d_{a}+i\sqrt{2\Pi^{+}}S_{a}.
\eq 
The next step consist in the deformation of the other first class constraints
$d_{\dot a}=0$, $T=0$ 
to be consistent with the new $\tilde d_{a}$ constraints. Consider the case of
$d_{\dot a}$. We need to find a solution to the condition
\be
\{\tilde d_{a}, \tilde d_{\dot a}\}=0.
\eq
The solution has the general form 
\be
\tilde d_{\dot a}=d_{\dot a}+A_{\dot a b}S_{b}+B_{\dot a [bc]}S_{b}S_{c},
\eq
where
\be
A_{\dot a b}=\frac{2i\gamma^{i}_{\dot a a}\Pi^{i}}{\sqrt{2\Pi^{+}}},\quad
B_{\dot a [ac]}=\frac{2\gamma_{\dot b[a}^{i}\gamma_{c]\dot a}^{i}\theta'_{\dot
b}}{\Pi^{+}}.
\eq
The extended constraint is
\be
\label{new-c-2}
\tilde d_{\dot a}=d_{\dot a}+ \frac{2i\Pi^{i}}{\sqrt{2\Pi^{+}}}(\gamma^{i}
S)_{\dot a}+\frac{2(\theta'\gamma^{i}S)(\gamma^{i}S)_{\dot a}}{\Pi^{+}},
\eq
keeping in mind that the last term has to be  antisymmetrized in undoted spinorial
indices. 

Now to find the extended constraint associated with $T$ it is easy to
proceed first to the extension of $\Pi_{\mu}$ defined in eq (\ref{Pi-mu}). To do
that we need to solve the condition (\ref{BFT-cond})  for $\tilde\Pi_{\mu}$, i.e.,
\be
\{\tilde d_{a}, \tilde \Pi_{\mu}\}=0.
\eq
A simple check shows that the series in powers of the new variables $S$ for
$\tilde \Pi_{\mu}$ stops up to second order terms. The solution is
\be
\tilde
\Pi^{\mu}=\Pi^{\mu}+4i\frac{(\theta'\gamma^{\mu}S)}{\sqrt{2\Pi^{+}}}+i\frac{S\gamma^{\mu}
S}{\Pi^{+}}.
\eq
Using this solution we will define the new first class constraint $\tilde T$ in such a way that it will
close in a Lie algebra with the rest of the new constraints 
\be
\label{new-c-3}
\tilde T=\frac{\tilde
\Pi^{2}}{4\Pi^{+}}=-\frac{\Pi^{-}}{4}+\frac{\Pi^{i}\Pi^{i}}{4\Pi^{+}}+
2i\frac{\theta'_{a}S_{a}}{\sqrt{2\Pi^{+}}}+i\frac{S_{a}S'_{a}}{2\Pi^{+}}+
4i\frac{\Pi^{i}(\theta'\gamma^{i}S)}{(2\Pi^{+})^{3/2}}-2\frac{(\theta'\gamma
S)^{2}}{(\Pi^{+})^{2}}.
\eq
The new effective gauge algebra in the extended space is now
\be
\{\tilde T,\tilde T\}=0,\quad \{\tilde d_{a},\tilde d_{b}\}=0,\quad \{\tilde
d_{\dot a},\tilde d_{a}\}=0,
\eq
\be
 \{\tilde d_{a},\tilde T\}=0, \quad \{\tilde
d_{\dot a},\tilde d_{\dot b}\}=-8i\tilde T \delta_{\dot a\dot b}. 
\eq 
The gauged GS first order action is
\be
\label{BFT-action}
\tilde S=-\frac12\int d^2\zeta\Big( \dot
x^{\mu}p_{\mu}+\dot\theta^{A}_{\alpha}p^{A}_{\alpha}+\frac{i}{2} \dot S^{A}_{a}
S^{A}_{a}-\lambda \tilde T-\hat\lambda\hat{\tilde T}-
\lambda^{A}_{\alpha}\tilde d^{A}_{\alpha}\Big),
\eq
where we have included the two sectors. Its gauge symmetries are the
worldsheet diffeomorphisms that are generated by $\tilde T$ and $\hat{\tilde T}$ and a
new fermionic gauge symmetry that is generated by $\tilde d^{A}_{\alpha}$. Of course
the theory is not manifestly Lorentz covariant but the Lorentz invariance is
guaranteed up to a BRST trivial transformation \cite{BM}.

We have now 17 first class constraints by sector and as expected the model is
equivalent to the original GS action and by construction has the same number of degrees 
of freedom.
Two comments are in order. The first is that this embedding of the GS
superstring is equivalent  to the classical  BM action. Its quantization can be
performed along the same lines as the quantization of the BM model. Subtitles
related to ordering ambiguities must be taken into account for a consistent
quantization of this action. Secondly, as the BM model can be related to the
pure spinor formalism via similarity transformations between the associated BRST
charges, this model can also be related to the pure spinor formalism using the same 
sequence of similarity transformations between its associated BRST charges. 
The advantage of our perspective
is that we have developed a gauge model in a completely systematic way 
starting from the plain GS
superstring and consequently we have more control over any change in the
embedding procedure that can be of help to relate GS and pure spinor
formalisms in a more direct way. This procedure can also be of some help to
better understand many aspects of pure spinor formalism like its geometrical
interpretation, the path integral measure, or the underlying action.

From the other hand our new constraints (\ref{new-c-1},\ref{new-c-2},\ref{new-c-3}) 
are the same as the ones obtained
in \cite{AK}. It is quite surprising for us that the constraints
are exactly the same. The two procedures are very
different. In \cite{AK} the number of fermions is doubled and an interaction between
them was introduced by hand. After fixing the semi-lightcone  gauge and making a
complicated Darboux transformation simplifying  the Dirac bracket, the results
of \cite{AK} coincides with the BFT embedding presented here. We will try to explain this 
relation in
the next section
by extracting more information about how the BFT embedding works.
 
\section{Relation with the AK model}
 
 That the embedding approach to GS action has something to do with the
 interacting action proposed in \cite{AK} is at first sight very surprising. Here we will  try 
 to elaborate on this relation using a slightly modified approach to the conversion 
 procedure. The arguments presented in this section does not apply to the case of a general 
 constrained system
 but are valid in some special type of systems like the one considered here.   
 
 Lets start by noticing that another way to apply the BFT embedding  
 is to seek for new extended
coordinates $\tilde x^{\mu},\tilde p_{\mu}, \tilde\theta^{A}_{\alpha},
\tilde p^{A}_{\alpha}$ such that they satisfy the conditions
\be
\label{cond-z}
\{\tilde d^{A}_{a}, \tilde z\}=0,
\eq
where $\tilde z(x^{\mu},p_{\mu},\theta_{\alpha},p_{\alpha},S)$ is any of 
the phase space extended new coordinates or momenta. If we can solve these
conditions then we can use the solutions to extend any observable of the
original phase space to the new extended phase space. The procedure is as
follows: suppose that we have a function in the original phase space $A(z)$.
First we write it as $A(\tilde z)$, then substitute $\tilde z$ by the
solution to (\ref{cond-z}) and find $\tilde A(z,S)$. For the GS superstring the solution
to the conditions (\ref{cond-z}) are\footnote{This solutions are for the sector 1. The
solution for the other sector has the same form.}
\be
\label{trans-x-theta}
\tilde x^{\mu}=x^{\mu}-i\frac{(\theta\gamma^{\mu}S)}{\sqrt{2\Pi^{+}}}, \quad
\tilde\theta_{a}=\theta_{a}-\frac{S_{a}}{\sqrt{2\Pi^{+}}},\quad 
\tilde\theta_{\dot a}=\theta_{\dot a},
\eq
for configuration space variables. For the bosonic momenta $p_{\mu}$ and the fermionic momenta 
$p_{\alpha}$ the solutions are
\be
\label{trans-p-mu}
\tilde p^{\mu}=p^{\mu}+i\Big(\frac{\theta\gamma^{\mu}S}{\sqrt{2\Pi^{+}}}\Big)',
 \eq
\be
\label{trans-p-a}
\tilde p_{\alpha}=p_{\alpha}-i\frac{(\gamma^{\mu}
S)_{\alpha}}{\sqrt{2\Pi^{+}}}(\Pi_{\mu}-W_{1\mu}+P_{\mu})+
i(\gamma^{\mu}\theta)_{\alpha}P_{\mu}+C_{\alpha},
\eq
where
\be
P_{\mu}=2i\frac{\theta'\gamma_{\mu} S}{\sqrt{2\Pi^{+}}}+
i\frac{S\gamma_{\mu} S'}{2\Pi^{+}}+i\Big(\frac{\theta\gamma_{\mu}
S}{\sqrt{2\Pi^{+}}}\Big)',
\eq
and
\be
C_{a}=i\sqrt{2\Pi^{+}}S_{a},\quad C_{\dot a}= \frac{2i\Pi^{i}}{\sqrt{2\Pi^{+}}}(\gamma^{i}
S)_{\dot a}+\frac{2(\theta'\gamma^{i}S)(\gamma^{i}S)_{\dot a}}{\Pi^{+}},
\eq
where the last term must be antisymmetrized with repect to the  
spinorial indices without dots. The new momenta $\tilde p_{\mu}$, $\tilde
p_{\alpha}$ in (\ref{trans-p-mu}, \ref{trans-p-a}) are very similar to the Darboux transformation
proposed in \cite{AK} to simplify the Dirac bracket but there are differences 
that we will explain below. What is perhaps
more interesting is that the transformations (\ref{trans-x-theta}) 
 in configuration space can be used to
obtain, from the GS action (\ref{GS-action}), the interacting AK action. Indeed,
redefine $S_{a}/\sqrt{2\Pi^{+}}$ as $\xi_{a}$ and use (\ref{trans-x-theta}) to get 
\be
\label{AK-action}
S=-\frac12 \int d^{2}\zeta \Big[\sqrt{-g}g^{ij}\Pi^{\mu}_{i}\Pi_{\mu
j}+2\varepsilon^{ij}\Pi^{\mu}_{i}(W^{1}_{j\mu}-W^{2}_{{j\mu}})-2
\varepsilon^{ij} W^{1\mu}_{i}W^{2}_{j\mu}\Big],
\eq
where
\be
W^{A\mu}_{i}=i\Theta^{A}\gamma^{\mu}\partial_{i}\Theta^{A}, \quad 
\Pi^{\mu}_{i}=\partial_{i}x^{\mu}-\sum_{A}W^{A\mu}_{i}
-i\sum_{A}\partial_{i}(\theta^{A}\gamma^{\mu}{\xi^{A}}),
\eq
with 
$\Theta^{A}=\theta^{A}-\xi^{A}$ as in \cite{AK}\footnote{We denote by $\xi$ the
variable that are denoted as $\tilde\theta$ in \cite{AK} to avoid some possible confusion
with our tilde variables.}. Notice that the effect of the substitution of the new configuration variables
(\ref{trans-x-theta}) in terms of the old ones in the original GS action (\ref{GS-action}) 
produces a deformation of the symplectic structure and a deformation of the original
bosonic constraints. The first order action has now the form
\be
S=-\frac12\int d^2\zeta\Big( \dot
x^{\mu}p_{\mu}+\dot\theta^{A}_{\alpha}\Delta^{A}_{\alpha}+\dot
\xi^{A}_{a}\Xi^{A}_{a}-\lambda  \tau-\hat\lambda\hat{ \tau}\Big),
\eq
 where space-time momenta is
 \be
p_{\mu}=\Pi_{0\mu}-(W^{1}_{1\mu}-W^{2}_{1\mu}),
\eq
 and the functions in the kinetic term are
 \be
 \Delta^{1}_{\alpha}=-i(\gamma^{\mu}\xi^{1})_{\alpha}p_{\mu}+
 i(p^{\mu}-\Pi_{1}^{\mu}-W_{1}^{2\mu})(\Theta^{1}\gamma_{\mu})_{\alpha}, 
 \eq
 \be
 \Delta^{2}_{\alpha}=-i(\gamma^{\mu}\xi^{2})_{\alpha}p_{\mu}+
 i(p^{\mu}+\Pi_{1}^{\mu}+W_{1}^{1\mu})(\Theta^{2}\gamma_{\mu})_{\alpha},
 \eq
 and
 \be
 \Xi^{1}_{a}=-i(\gamma^{\mu}\theta^{1})_{a}p_{\mu}+
 i(p^{\mu}-\Pi_{1}^{\mu}-W_{1}^{2\mu})(\Theta^{1}\gamma_{\mu})_{a}, 
 \eq
 \be
 \Xi^{2}_{a}=-i(\gamma^{\mu}\theta^{2})_{a}p_{\mu}+
 i(p^{\mu}+\Pi_{1}^{\mu}+W_{1}^{1\mu})(\Theta^{2}\gamma_{\mu})_{a}.
 \eq
 $\tau,\hat \tau$ are the deformed bosonic constraints after the substitution of 
 (\ref{trans-x-theta}) in the original bosonic constraints $T$ and $\hat T$.
 Defining, as usual, the fermionic momenta as the coefficient that multiply $\dot\theta$ 
 in (\ref{AK-action}), we find the fermionic constraints 
 \be
 D^{A}_{\alpha}=p^{A}_{\alpha}-\Delta^{A}_{\alpha},
 \eq
  that correspond to the constraints $d^{A}_{\alpha}$ of the original GS action. Integration by
  parts in the kinetic term produces the action
 \be
S=-\frac12\int d^2\zeta\Big( \dot
x^{\mu}(p_{\mu}-P_{\mu})+\dot\theta^{A}_{\alpha}(p^{A}_{\alpha}-P^{A}_{\alpha})
+i\Pi^{+}\dot\xi^{A}_{a}\xi^{A}_{a}-\lambda  \tau-\hat\lambda\hat{ \tau}-\lambda^{A}_{\alpha} 
D^{A}_{\alpha}\Big),
\eq
where 
\be
P_{\mu}=-i(\theta\gamma^{\mu}\xi)'_{1}+i(\theta\gamma^{\mu}\xi)'_{2},
 \eq
 and $()_{1}$ denotes variables of the sector 1 and $,()_{2}$ for sector 2. 
 The fermionic momenta redefinition is
 \be
 P_{a}=i{x'}^{+}\xi_{a}+R_{a},\quad P_{\dot a}=(\gamma^{i}\xi)_{\dot a}\Big(i x'_{i}+
 (\theta\gamma_{i}\xi)_{2}\Big)+R_{\dot a} , 
 \eq
 where
 \be
 R_{\alpha}= (\theta \gamma^{\mu})_{\alpha}\Big(-2(\xi\gamma_{\mu}\theta')+
 \xi\gamma_{\mu}\xi'
 +(\theta\gamma_{\mu}\xi)'\Big)+(\xi\gamma^{\mu})_{\alpha}
 \Big(3(\theta\gamma_{\mu}\theta')-2(\theta\gamma_{\mu}\xi')\Big).
 \eq
The redefinition of the space time momenta are 
the same as the redefinition that we have used in the BFT embedding but the redefinitions associated 
with the fermionic momenta are slightly different. The reason is that the fermionic constraints
$D^{A}_{\alpha}$ are not the same as the original fermionic constraints $d_{\alpha}$ {\em after} the
substitution of the new coordinates (\ref{trans-x-theta}). Now it is easy to check that the
field redefinitions 
\be
p_{\mu}\to p_{\mu}-P_{\mu}, \quad p^{A}_{\alpha}\to p^{A}_{\alpha}- P_{\alpha},
\quad  \xi_{a}=S_{a}/\sqrt{2\Pi^{+}}, 
\eq
produces the first order action (\ref{BFT-action}) obtained by the BFT embedding. The efficient 
way, just presented to 
analyze the constraint structure of the action (\ref{AK-action}) is inspired in the 
Faddeev-Jackiw method that
is equivalent to the Dirac method for a wide class of constrained systems \cite{FJ}.    So
we have obtained the interacting model \cite{AK} from the BFT embedding approach using
the configuration variables $\tilde x^{\mu}, \tilde\theta^{A}_{\alpha}$ obtained by solving
the embedding condition (\ref{cond-z}). Notice that this procedure does not work for a 
general constrained system. The fact that the GS action is linear in the velocities of the fermionic 
variables and that the solution to (\ref{cond-z}) depends only on the configuration variables,
after the redefinition that relates $S$ with $\xi$ are crucial ingredients in the construction of
a Lagrangian action compatible with the dynamics of the original Hamiltonian action.

\section{Conclusions}

We have worked out the BFT embedding of the GS formalism using light-cone
variables. We have shown that the correction to the second class constraints
and the embedding of the first class algebra requires only terms up to second
order in the new fermionic variables and its derivatives. The
embedding can be performed in a completely systematic way and we do not have
to fix the gauge at any stage of our embedding procedure not to add ad-hoc
variables other than the ones needed by the BFT embedding approach. The gauged
GS action that results from our analysis is equivalent to the BM model
proposed recently in \cite{BM}. It also explain some aspects of the rationale under the
construction of a related model worked in \cite{AK}.

The systematics behind our approach can be used to study the relation of
the BRST charges and associated  cohomologies between this model and the pure
spinor formalism. We also have more control on the gauge fixing procedure. 
It is also a better point
of departure to study the supermembrane and other topics not yet well understood
in the pure spinor formalism, like the associated action and the measure of
the path integral. It could  be also of interest to explore the idea of 
non-Abelian \cite{BGL} conversion to simplify the relation between the Berkovits pure 
spinor formalism and the GS embedding. We will return to these aspects elsewhere.

 \section*{Acknowledgements}

This work was partially supported by Mexico's
National Council of Science and Technology (CONACyT), under grant
CONACyT-40745-F, and by DGAPA-UNAM, under grant IN104503-3.

 \section*{Appendix}
 
The notation used in the paper is the following: for the 
GS action (\ref{GS-action}), $\varepsilon^{01}=1$, $\theta^{A}_{\alpha}$, 
is an SO(9,1) spinor with $A=1,2$  
supersymmetries and 
$\alpha=1,2,...16$ components. They are real Mayorana-Weyl spinors of the same chirality. 
$x^{\mu}$
$\mu=0,1,2...9$ are the space-time configuration variables. The worldsheet coordinates 
are $\zeta=(t,\sigma)$ and the derivatives with respect to time will be denoted by a dot and
with respect to sigma by a prime. The $\gamma^{\mu}$ matrices are $16\times 16$ 
Dirac matrices, real and symmetric. The convention for derivatives are left derivatives 
and then
\be
\{\theta^{A}_{\alpha},p^{B}_{\beta}\}=-\delta_{\alpha\beta}\delta^{AB},
\eq
This convention fixes the order of $\dot\theta$ and $p$ in the kinetic term of the 
first order action. 

Light-cone coordinates: We split the spinorial index $\alpha$ according to the 
the SO(8) chiral and anti-chiral components $a$ and $\dot a$. Space-time indices 
decompose according to $\gamma^{\pm}=\gamma^{0}\pm \gamma^{9}$, 
$x^{\pm}=x^{0}\pm x^{9}$... and $\gamma^{i}$, $x^{i}$, i=1,2,...8 for the other 
set of vector components. The Dirac algebra decomposes according to
\be
\gamma^{+}_{\dot a\dot b}=-2\delta_{\dot a\dot b}, \quad 
 \gamma^{-}_{ab}=-2\delta_{ab}, \quad 
 \gamma^{i}_{\dot a a}\gamma^{j}_{\dot a b}+ \gamma^{j}_{a\dot a }
 \gamma^{i}_{\dot a b}=2\delta^{ij}\delta_{ab},
 \eq
 and 
\be
 \gamma^{i}_{a \dot b}\gamma^{i}_{c \dot d}+ \gamma^{i}_{a\dot d }
 \gamma^{i}_{c \dot b}=2\delta^{ac}\delta_{\dot b\dot d},
\eq
with $\gamma_{\dot a a}$ symmetric. We use repeatedly through the text 
the Fierz identity
\be
(\gamma_{(\alpha\beta})^{\mu}(\gamma_{\gamma)\delta})_{\mu}=0.
\eq

 \end{document}